\documentclass[10pt,aps,showpacs,twocolumn,superscriptaddress,natbib]{revtex4}
\usepackage{graphicx}
\usepackage{dcolumn}
\usepackage{amsmath}

\begin{document}

\title{Characterization of A Novel Avalanche Photodiode for Single Photon Detection in VIS-NIR range}

\begin{abstract}
\rule{0ex}{3ex}
\noindent
{\bf
In this work we investigate operation in the Geiger mode of the new single photon avalanche photo diode (SPAD) SAP500 manufactured by Laser Components. This SPAD is sensitive in the range 400-1000nm and has a conventional reach-through structure which ensures good quantum efficiency at the long end of the spectrum. By use of passive and  active quenching schemes we investigate detection efficiency, timing jitter, dark counts, afterpulsing, gain and other important parameters and compare them to the "standard" low noise SPAD C30902SH from Perkin Elmer. We conclude that SAP500 offers better combination of detection efficiency, low noise and timing precision.
}
\end{abstract}

\author{M. Stip\v cevi\' c}
\email{Mario.Stipcevic@irb.hr}
\affiliation{\footnotesize Rudjer Bo\v{s}kovi\'{c} Institute,
         Bijeni\v cka 54, P.O.B. 180, HR-10002 Zagreb, Croatia}
\author{H. Skenderovi\' c}
\affiliation{\footnotesize Institute of Physics,
         Bijeni\v cka XX, HR-10002 Zagreb, Croatia}
\author{D. Gracin}
\affiliation{\footnotesize Rudjer Bo\v{s}kovi\'{c} Institute,
         Bijeni\v cka 54, P.O.B. 180, HR-10002 Zagreb, Croatia}

\maketitle

\section{Introduction}

Experiments in quantum information and communication are mostly concentrated on studying photonic state manipulation at the quantum level. This is because photons, among all elementary particles, have unique properties of being easily produced in abundance, easily manipulated, easily transmitted to large distances and yet being relatively easy to detect. 
At the current state of the art main tools for experimental research in quantum information and communication (lasers, fiber optic light guides, sources of entangled photons ...) require single photon detection in the near infrared range of 700-1550nm \cite{downqkd1,downqkd2,downqkd3,downqkd4,weiunfuterrng,qrbg}.

Photomultipliers, as traditional photon detectors, have very small quantum efficiency and a large noise in that wavelength range. The next most mature technology of detecting single photons is based upon avalanche photo diodes (APD) \cite{dautet}. Not all APDs are suitable for that task. Special semiconductor structures, so called SPAD (Single Photon Avalanche Diode), exhibit characteristics required for single photon detection \cite{Cova2004}. With respect to photomultipliers silicon SPADs offer superior quantum efficiency in the range 500-1000nm, higher gain, mechanical robustness, possibility of miniaturization, low power consumption and relatively low cost.

Since early 1980's practically the only commercially available silicon APD suitable for single photon detection was the C30902SH from PerkinElmer. Single photon sensitivity is reached in the Geiger mode where a single photoelectron may trigger an avalanche pulse of about 10$^8$ carriers \cite{c30902-datasheet}.    
In this work we have tested a brand new SPAD SAP500 from Laser Components and compared it to performance of the "standard" C30902SH, in the Geiger mode. Both photodiodes utilize the reach-through type of structure, have the same photo sensitive diameter of 0.5mm and similar spectral sensitivity range.

\section{Quenching circuits}

Single photon detection by an APD assumes reverse bias voltage $V_R$ greater than the "Geiger" or "breakdown" voltage $V_{BR}$. We define "overvoltage" as $V_{over}=V_R - V_{BR}$.

\begin{figure}[h]
\centerline{\includegraphics[width=85 mm,angle=0]{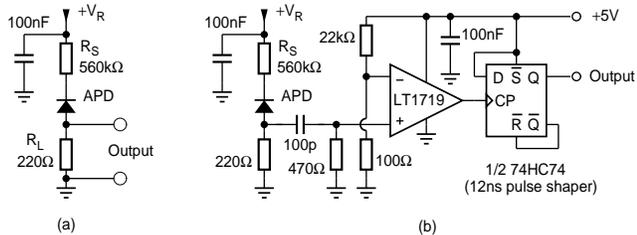}}
\caption{ (a) the passive quenching circuit; (b) the same followed by the constant level discriminator and the pulse shaper.} 
\label{passives}
\end{figure}

In order to fully assess characteristics of APD's in Geiger mode one must use some kind of a quenching circuit. In most measurements we have used a simple (but effective !) passive quenching (PQ) circuit shown in Fig. \ref{passives}(a), optionally followed by a home-made constant level discriminator (CLD) and a 12ns pulse shaper as shown Fig. \ref{passives}(b). A threshold level of the CLD is fixed at 22.5mV in both circuits. If the current limiting resistor $R_S$ is chosen sufficiently large, an avalanche will cease (quench) by itself within a sub-nanosecond time. 
A condition for successful quench is that $V_{over}/R_S$ is smaller than the {\em latch current} of the given SPAD. 
In that case, the avalanche current will quickly discharge junction and parasitic capacitances until voltage accross the SPAD drops below $V_{BR}$ causing the current to drop below the latch current after which the avalanche will quench.
After a quench, voltage across the diode recovers towards its initial value following the exponential law with the time constant $\tau_R = R_S (C_{SPAD}+C_p) \approx 0.9\mu s$, where $C_{SPAD}$ is the capacitance of the reversely polarized SPAD and $C_p$ is the parasitic capacitance present in the actual circuit (the capacitances will be measured in the section \ref{response-cap}). During the voltage recovery, the detection efficiency of the SPAD changes from zero towards the initial value. 
Due to the capacitive coupling, the circuit in Fig. \ref{passives}(b) can measure poissonian time-distributed avalanches up to over 2MHz and is highly immune to paralyzation effect \cite{parart}. 

A very useful characteristic of the PQ circuit is that it allowed us to operate SPADs at an overvoltage of up to 30V. However, 
for those measurements for which the long recovery time of the PQ circuit was prohibitive we have used our previously described active quenching (AQ) circuit \cite{aqmario2007}. 
In active quenching, both lowering and restoring of bias voltage are assisted by active electronic components enabling a quick recovery and well defined dead time. Neglecting the transition times, actively quenched SPAD is either completely insensitive to incoming photons or is at its nominal sensitivity. Due to inevitable delays in the electronics, the dead time of an AQ circuit is somewhat longer than the quenching time $t_Q$ during which the SPAD is actually kept below $V_{BR}$. In our case $t_{dead} \approx t_Q + 20$ns.
Slight variations of a single capacitor allowed us to choose the dead time in the interval of 25-50ns.

\section{Single photon response and effective capacitance}
\label{response-cap}

\begin{figure}[h]
\centerline{\includegraphics[width=80 mm,angle=0]{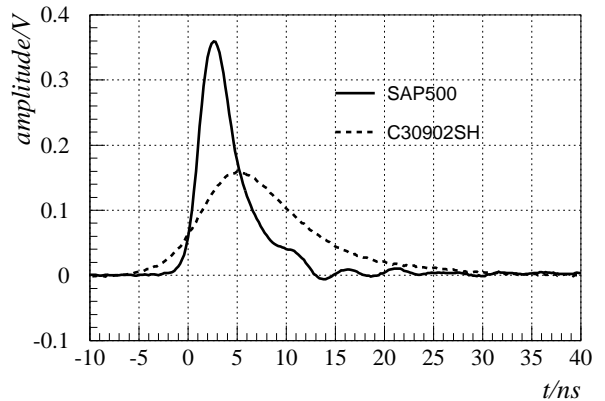}}
\caption{A single photon response at the output of the passive quenching circuit, for SAP500 and C30902SH. Because of pulse to pulse variations, a statistical average of 256 output waveforms are shown.} 
\label{pulses}
\end{figure}

Figure \ref{pulses} shows average single photon responses of the two SPADs as measured at the output of the PQ circuit (Fig. \ref{passives}(a))

\begin{figure}[h]
\centerline{\includegraphics[width=80 mm,angle=0]{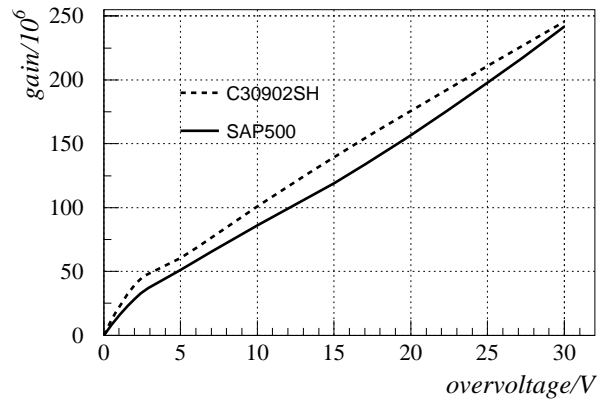}}
\caption{ Multiplication gain (in millions) as a function of overvoltage for single photons of 676nm at room temperature.} 
\label{gain}
\end{figure}

at $V_{over}=5$V. We define multiplication gain $G$ as a number of electrons produced (on average) by a single converted photon:
\begin{equation}
G = \frac{1}{R_L e} \int_0^\infty V(t)dt .
\label{Gain}
\end{equation}
where V(t) is the voltage at the output of the PQ circuit.  
Gain as a function of overvoltage for single photons of 676nm is shown in Fig. \ref{gain}. Gains of both diodes are of the similar magnitude but due to its much shorter pulse SAP500 produces a higher peak. Interestingly, gain of these tiny devices is up to a couple of hundred times higher than the gain of conventional 10 dynode photo multipliers which is typically on the order of $10^6$.

By virtue of charge conservation, the total charge delivered by the avalanche is equal to the product of the SPAD's capacitance (plus any parasitic parallel capacitance) and the voltage step the SPAD makes during the quench. If a capacitance $C_{||}$ is added in parallel to the SPAD, the charge delivered within one pulse rises by the factor
$(C_{SPAD} + C_p + C_{||}) / (C_{SPAD} + C_p)$ where $C_p$ is the parasitic capacitance estimated to be 1.10pF.  
Using $C_{||}=$2.7pF we obtained capacitances of 1.50pF for C30902SH and 1.63pF for SAP500.
Although the two capacitances are very similar in magnitude, SAP500 produces narrower and higher pulse with roughly 2.9 times faster rise time (10-90\%) than C30902SH (Fig. \ref{pulses}) leading to a much better time resolution, as will be established by direct measurements. Faster discharge of the junction plus parasitic capacitance by SAP500 could be attributed to its smaller space-charge resistance \cite{quiang1993} and therefore larger avalanche current. 

\section{Dark counts}

\begin{figure}[h]
\centerline{\includegraphics[width=80 mm,angle=0]{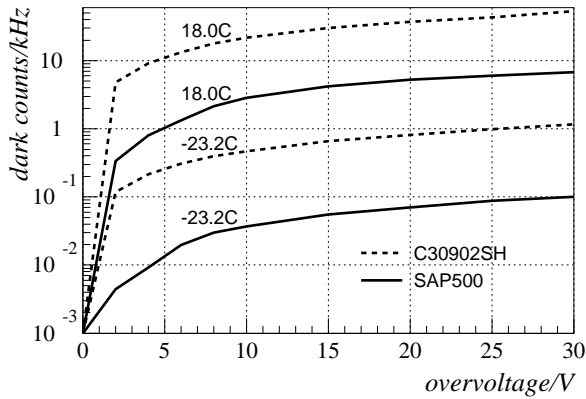}}
\caption{Dark counts rate as function of overvoltage measured at two temperatures.} 
\label{darkcounts}
\end{figure}

In photon counting technique dark counts present an unwanted noise. The highest tolerable noise in most applications lies between a few tens Hz and about 1kHz. For such a low average counting frequency, performance of a SPAD can be assessed quite precisely with the passive quenching circuit, which allows us to check the dark counts rate at an overvoltage up to 30V. Fig. \ref{darkcounts} shows dark counts rate for C30902SH and SAP500 at 18.0$^{\rm o}$C and -23.2$^{\rm o}$C as a function of overvotage. 

We can see that cooling from 18.0$^{\rm o}$C to -23.2$^{\rm o}$C reduces dark counts rate more for SAP500 than for C30902SH. Reduction factor for C30290SH is about 46 while for SAP500 it is about 75.

While the C30902SH has been factory selected for low noise, the above tested SAP500 sample has been randomly chosen from a batch. In order to check the consistency of this measurement we have tested further 6 samples which have been selected such that the $V_{BR}$ was between 115 and 125V at 18$^{\rm o}$C. The dark counts of samples were spread between 1.7kHz and 3.5kHz whereas the tested sample had 2.2kHz, all measured  at 18$^{\rm o}$C and $V_{OVER}$=8V.

\section{Afterpulsing probability}
\label{apsection}

\begin{figure}[h]
\centerline{\includegraphics[width=80 mm,angle=0]{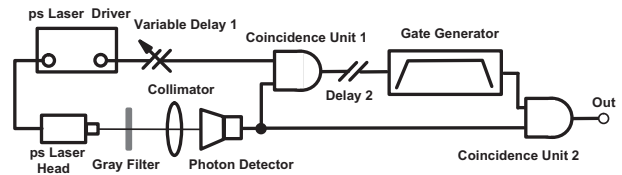}}
\caption{Setup for measurement of afterpulsing probability.}
\label{apsetup}
\end{figure}

Afterpulse in a SPAD is caused by a carrier left over from a previous avalanche, trapped in an impurity and then released at a latter time \cite{cova-afterp03}.
If such a carrier makes its way to the avalanche region it may cause an avalanche which is indistinguishable from a true photon detection.
Important parameters for afterpulsing probability are impurity concentration and carrier lifetime. 
In silicon SPADs afterpulsing is a fast process decaying in about few tens to few hundreds nanoseconds therefore the active quenching is necessary in order to capture afterpulses and measure the afterpulsing probability. 
We define afterpulsing probability (a.p.) as a probability that an afterpulse will appear after a detection of a photon. According to this definition we have constructed a setup for measuring the a.p. as shown in Fig. \ref{apsetup}.

Ultra short, weak pulses of light from a pulsed laser (PicoQuant PDL 800-D, laser head 39ps FWHM, $\lambda$=676nm) are fired into the detector. Pulse energy and gray filter are chosen such that the detector receives almost exclusively either 0 or 1 photon per pulse whereas two and more photon events are strongly suppressed. Delay 1 is adjusted such that in case that a photon has been fired and detected, the Coincidence Unit 1 generates a logic pulse. Due to a short coincidence window (20ns) and low dark counts rate (up to a few kHz) accidental coincidences are negligible. The pulse is delayed for 100ns (Delay 2) after which the Gate Generator opens the gate of the Coincidence Unit 2 for the next 400ns. 

\begin{figure}[h]
\centerline{\includegraphics[width=80 mm,angle=0]{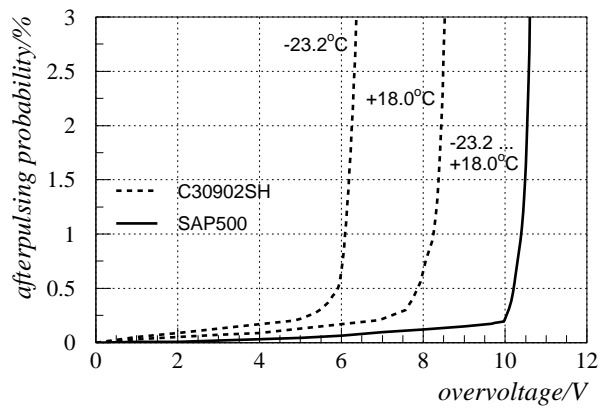}}
\caption{Afterpulsing probability as a function of overvoltage for the two SPADs, at various temperatures, measured by active quenching.} 
\label{apvsover}
\end{figure} 
 
In such a way afterpulses generated between 100-th and 500-th nanosecond after the photon detection appear at the output. 
The afterpulsing probability is then simply given as the ratio of
the frequency of afterpulses at the output of the second coincidence unit and  
the frequency of detected pulses at the output of the first coincidence unit.
Measured afterpulsing probabilities as functions of overvoltage are shown in Fig. \ref{apvsover}. We see that both diodes perform well up to some critical overvoltage where a.p. rises quickly. The rise is the effect of the active quenching circuit as will be explained in the next section. 

\section{Undervoltage}
\label{undervoltage}

Avalanches in Geiger mode appear at an inverse bias voltage greater than $V_{BR}$. Successful active quenching of such avalanches requires lowering the bias voltage a few voltages below $V_{BR}$, namely to $\le V_{BR}-V_U$ where $V_U$ is the "undervoltage". This effect has already been reported for the C30902S in \cite{dautet} but undervoltage was not precisely measured. Further condition for successful quench is that the bias is kept low for long enough time to allow annihilation of carriers participating in the avalanche. As noted in section \ref{apsection}, a carrier may get trapped in an impurity and released at a latter time causing an afterpuls. The lifetime of a trapped carrier is inversely proportional to the absolute temperature of the SPAD. 

\begin{figure}[h]
\centerline{\includegraphics[width=80 mm,angle=0]{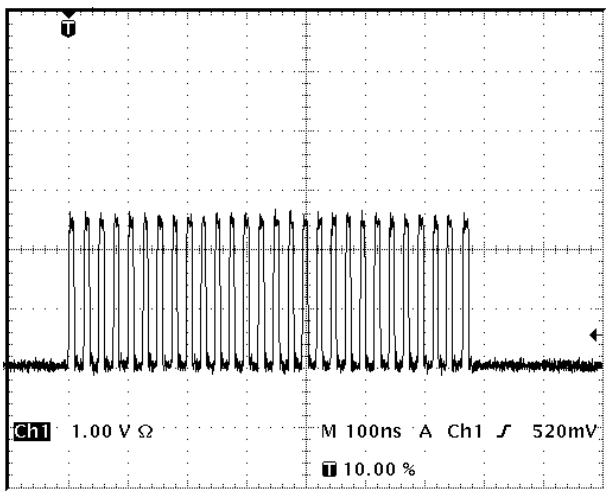}}
\caption{Inefficient active quenching condition: a cascade of quenching attempts.} 
\label{badquench}
\end{figure}

In this section we determine the undervoltages of both SAP500 and C30902SH. 
Our AQ circuit produces a fixed quenching voltage step of 12V. We have found that successful quenching at 18.0$^{\rm o}$C can be accomplished with $t_Q$=16ns for C30902SH and $t_Q$=9ns for SAP500.
While operation at 18.0$^{\rm o}$C is not really of practical importance because of the large noise, it is favourable for measurement of the undervoltge due to short lifetime of trapped carriers.
With temperature and $t_Q$ properly set up, measuring the $V_U$ consists of finding the highest value of overvoltage (the critical overvoltage, $V_C$) at which quenching can be done efficiently. If the overvoltage is set too high, there will be an elevated probability of unsuccessful quench after which the AQ circuit will initiate another quenching attempt and so forth until quenching finally succeeds. Looking at the output of the detector, this effect can be seen on the oscilloscope as a series of pulses separated by approximately one detector dead time (Fig. \ref{badquench}). 
 
Because our setup counts the additional quenching attempts as (fake) afterpulses, the critical overvoltage is characterized by a sharp rise in so measured afterpulsing probability \cite{aqmario2007}. This sharp rise can be seen in the measurements in Fig. \ref{apvsover}.
The undervoltage is now simply the difference between the voltage step provided by the AQ circuit and the critical overvoltage: $V_U=12V-V_C$.
From the Fig. \ref{apvsover} we estimate the undervoltages of C30902SH to be $\approx$4-4.5V and of SAP500 $\approx$2V. We also see that transition from efficient to inefficient quenching regime is much sharper for SAP500 than for C30902SH which offers more stable operation near the critical overvoltage.

In order to arrive to the dark counts levels of few hundred Hz or less, the operating temperature of both SPADs has to be significantly lowered.
As the temperature is lowered, trapped carriers lifetime becomes larger and more of them survive quenching process causing afterpulses. This effect is particularly strong for C30902SH which cannot be operated at overvoltage greater than 6V at -23.2$^{\rm o}$C even with $t_Q$ enlarged to 30ns (maximum permitted by our AQ circuit). 
On the other hand, the performance of SAP500 is virtually intact at -23.2$^{\rm o}$C and $t_Q$ of only 9ns. This extraordinary performance could be explained with either very short trap lifetime or by a low density of impurities in the avalanche region.
 
\section{Time resolution (jitter)}
\begin{figure}[h]
\centerline{\includegraphics[width=80 mm,angle=0]{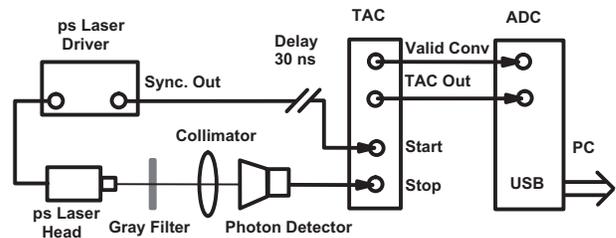}}
\caption{ Setup for measuring timing jitter of a SPAD based photon detector by means of a picosecond laser and time-to-amplitude converter (TAC).}
\label{jitter}
\end{figure}

In many experimental methods such as time resolved spectroscopy, quantum communication or range finding, precise timing of photon arrival is essential.
Photon arrival time resolution (timing jitter) has been measured by use of the passive quenching circuit with the constant level discriminator shown in Fig. \ref{passives}(b), and the whole measurement setup is shown in Fig. \ref{jitter}. 

\begin{figure}[h]
\centerline{\includegraphics[width=80 mm,angle=0]{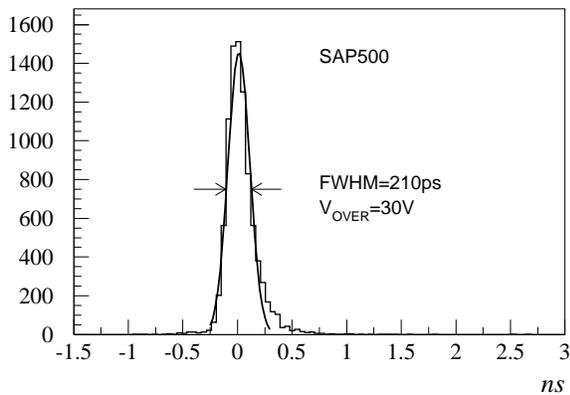}}
\caption{ Near Gaussian jitter distribution for SAP500 at -23.2$^{\rm o}$C and overvoltage of 30V. The right side tail, characteristic of reach-through structure, is barely visible.}
\label{jitter-distr}
\end{figure}

Arrival time is measured with a resolution of 10ps FWHM using the time-to-amplitude converter (TAC) Ortec model 567 whose analog output is read out by an 16 bit ADC (National Instruments, model NI USB-6251). Width of the laser pulse (39ps FWHM) and jitter of the Laser Driver (20ps FWHM) are negligible in these measurements.  The measured jitter is nearly a Gaussian function.  For illustration we show the distribution for SAP500 at $V_{over}=30$V and $T_{SPAD}=-23.2^{\rm o}$C (Fig. \ref{jitter-distr}). A fitted Gaussian is also shown. The right side tail, characteristic of reach-through structure, is barely present.

\begin{figure}[h]
\centerline{\includegraphics[width=80 mm,angle=0]{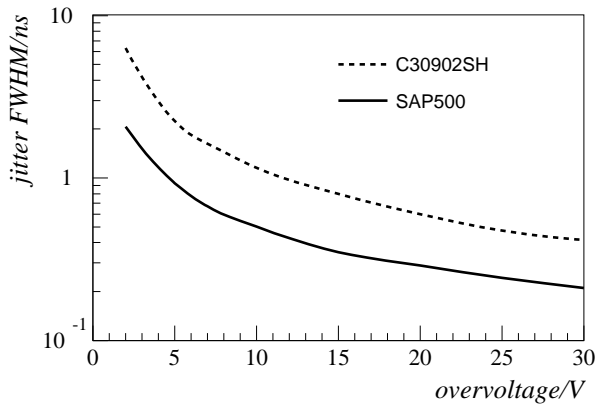}}
\caption{ Timing resolution (jitter) FWHM as a function of overvoltage for the two SPADs, measured with the passive quenching circuit and a constant fraction detector. At overvoltage of 30V jitter goes down to 420ps for C30902SH and 210ps for SAP500.}
\label{jitter-curves}
\end{figure}

Jitter as a function of overvoltage is shown in Fig. \ref{jitter-curves}. Measurements have been made at a low temperature (-23.2$^{\rm o}$C) at which SPADs are normally used. As is well known, SPADs utilizing reach-through structure are not optimal for timing purposes \cite{Cova2004}. Nevertheless, we see that SAP500 offers quite good timing resolution (450ps at $V_{over}$=10V, 210ps at $V_{over}$=30V), comparable to the best photomultiplier based single photon detectors, even with this simple passive circuit. A further improvement of resolution can be expected with use of a constant fraction detection principle \cite{quiang1993}. 

\section{Absolute detection efficiency at 810nm}

\begin{figure}[h]
\centerline{\includegraphics[width=80 mm,angle=0]{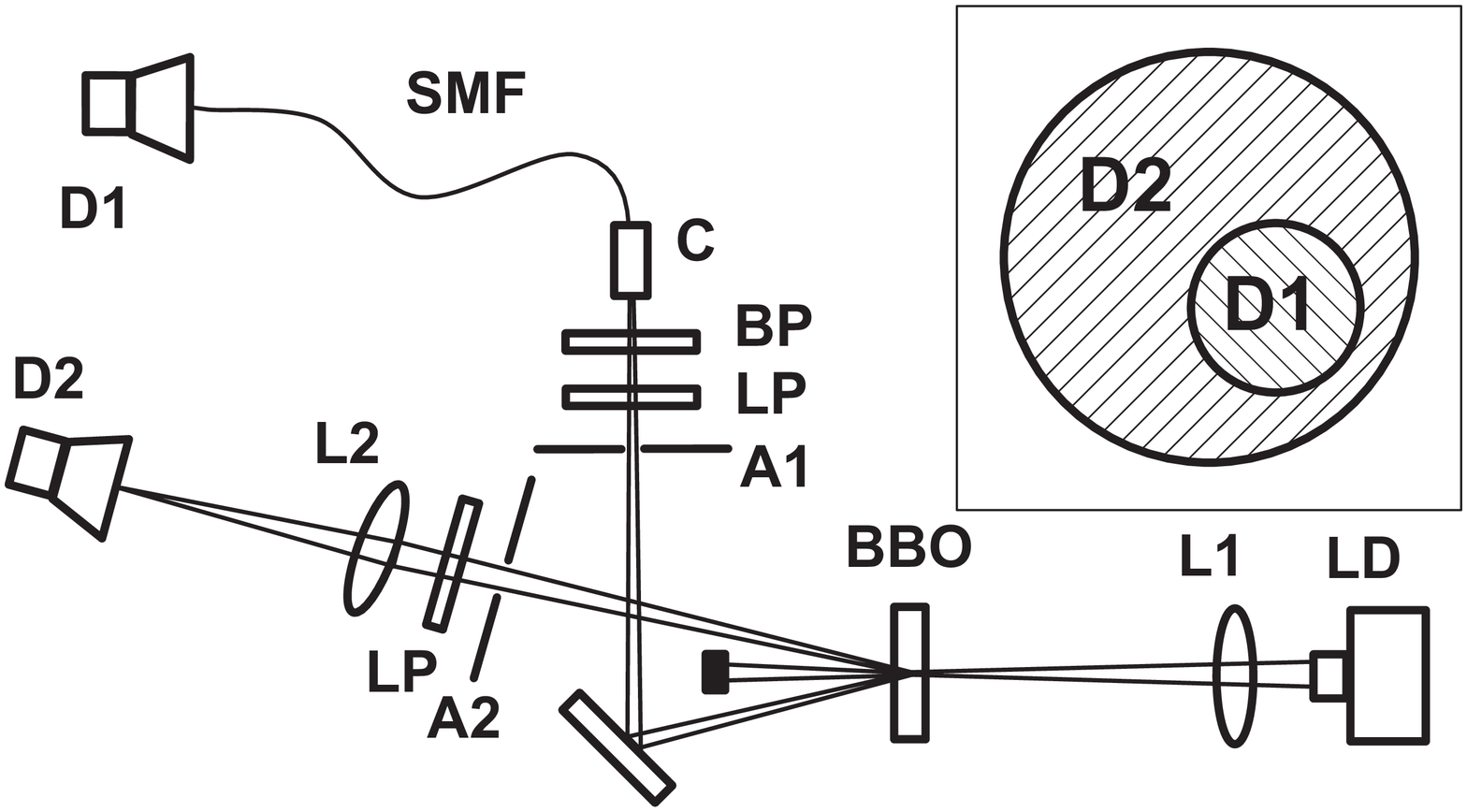}}
\caption{ Setup for measuring absolute detection efficiency at 810nm by use of the type-II parametric downconversion in BBO crystal.}
\label{desetup}
\end{figure}

Photon detection efficiency (PDE) is defined as the ratio of frequency of detected photons  and frequency of photons impinging the active surface of the APD.
Thanks to the development of the technique of spontaneous parametric downconversion, it is now possible to determine absolute photon detection efficiency of single photon detectors in a relatively simple manner using a method introduced in \cite{klyshko}. Having prepared the setup, measurement of absolute detection efficiency requires only a small number of measurements and no standards. Our setup, shown in Fig. \ref{desetup}, follows the general idea of \cite{rarity3}, which we explain here in a nutshell.    

One starts with two detectors of unknown efficiencies (D1 and D2) and wants to determine efficiency of the D2. A type-II spontaneous parametric downconversion setup, consisting of 405nm laser pump and conveniently cut BBO crystal, produces pairs of photons. Two photons in a pair are emitted almost simultaneously ($\Delta t\sim$100fs) each within a very thin cone. The two cones are tilted $\pm$3 degrees from the direction of the pump. One photon from a pair is directed towards D2. In this arm one wants transmission loses to be as small as possible (in order to minimize the uncertainty associated with these loses) therefore only a longpass filter LP (Thorlabs FGL715) needed for reducing the laser glow and anti-reflective coated collimating lens are used. The other photon from the pair is directed towards pigtailed auxiliary detector D1 (PerkinElmer SPCM-AQRH-14).

The crucial consideration is that the detector D1 must have a smaller viewing angle than D2 in such a way that if D1 has detected a photon then D2 has surely received (but not necessarily detected) the other photon from the pair. This condition can be achieved by adjustment of position and apertures A1 and A2. 
In the other arm, it is important that the spot of light collimated onto D2 fits well within the SPAD chip size (500$\mu$m dia.), otherwise the estimated efficiency would be lower than the true efficiency. In our case, spot size has been measured by a CCD camera to be only 20$\mu$m FHWM.
With all is in mind one has:

\begin{eqnarray}
F_1 & = & T_1 \epsilon_{1}^{'} + N_1 \\
F_2 & = & T_2 \epsilon_{2}^{'} + N_2 \\
C   & = & T_1 \epsilon_{1}^{'} \epsilon_{2}^{'} + N_1 F_2 \tau_c
\label{de-eqns}
\end{eqnarray}

where $T_i$, $F_i$, $N_i$ and $\epsilon_{i}^{'}$ are: true frequency of photons falling on the detector, frequency of detected photons, noise (dark counts) and effective detection efficiency which includes transmission loss, respectively for the i-th detector. $C$ is the coincidence rate within the time window of $\tau_c=$5.5ns.  
In our case, the coincidence rate is about C$\approx$3kHz while the term $N_1 F_2 \tau_c$ amounts only 3-4 Hz and is therefore negligible. After elimination of unknown $T_1 \epsilon_{1}{'}$ from the first and last line we obtain the expression for the efficiency of the detector D1:

\begin{equation}
\epsilon_2 = k\epsilon_{2}^{'} = k\frac{C}{F_1-N_1}.
\label{eq-eff}
\end{equation}

where the factor $k\approx 1.10$ accounts for transmission losses in the filter and lens in front of the D2. Since in our case detection frequency $F_1$ is quite low ($<$10kHz) dead time correction is not required. We use a coated lens with transmission coefficient of nearly 1. The transmission coefficient of the longpass filter LP was measured quite precisely as a factor of drop of detection rate of D2 when another identical filter is inserted.
Filters in the other arm have no effect to the measurement, except that the bandpass filter BP (810$\pm$5nm, Thorlabs FB810-10) defines the wavelength at which the PDE is measured.
Knowing the absolute detection efficiency at 810nm
allows us to properly scale  the relative spectral efficiency curve (measured below) so that it becomes an absolute one. 

\section{Spectral detection efficiency}
\begin{figure}[h]
\centerline{\includegraphics[width=65 mm,angle=0]{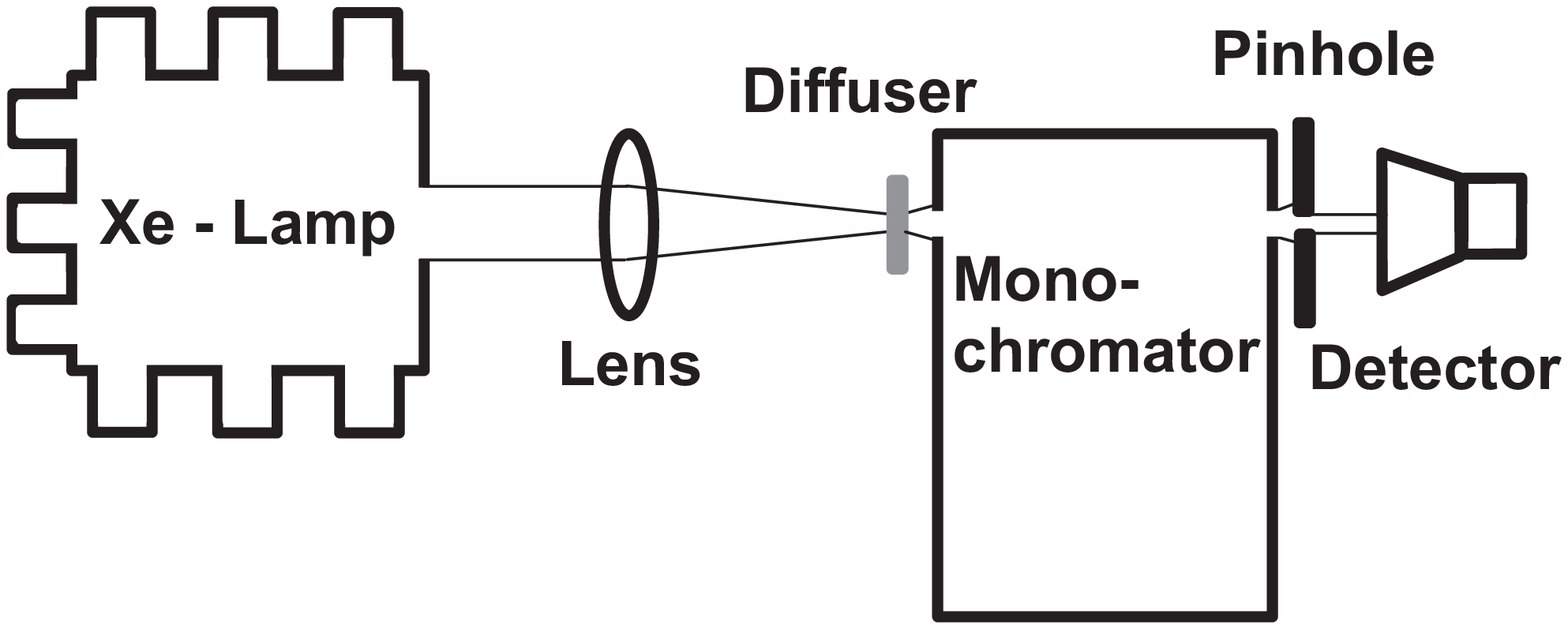}}
\caption{ Setup for measuring relative absolute detection efficiencies at various wavelengths by use of a monochromator.}
\label{spectral-setup}
\end{figure}

Relative spectral detection efficiency has been measured with the setup shown in Fig. \ref{spectral-setup}. 

Power $P_{emit}$ emitted from the output of the monochromator (Spectral Products CM110) at chosen wavelengths is known (calibrated). The detector under test is homogeneously illuminated through a pinhole by a small fraction of the emitted light power, $P_{inc}$. The photon detection efficiency $\epsilon$ at a given wavelength $\lambda$ is then given by:

\begin{equation}
\epsilon(\lambda) = \frac{f_{det}(\lambda)}{f_{inc}(\lambda)} = \frac{f_{det}(\lambda)}{P_{inc}(\lambda)/(hc/\lambda)} = k \frac{f_{det}(\lambda)}{\lambda P_{emit}(\lambda)}.
\label{spectral-eff}
\end{equation}
where $f_{det}$ is frequency of detected photons, $f_{inc}$ is frequency of photons impinging on the detector's sensitive area, and $k$ is a constant defined by the geometry of the setup. This constant can be determined by a single known value of detection efficiency at some wavelength, for example previously measured $\epsilon(810nm)$:

\begin{equation}
k = \frac{\epsilon(810{\rm nm}) \times 810{\rm nm} \times P_{emit}(810{\rm nm})}{f_{det}(810{\rm nm})}.
\label{k-spectral-eff}
\end{equation}

\begin{figure}[h]
\centerline{\includegraphics[width=80 mm,angle=0]{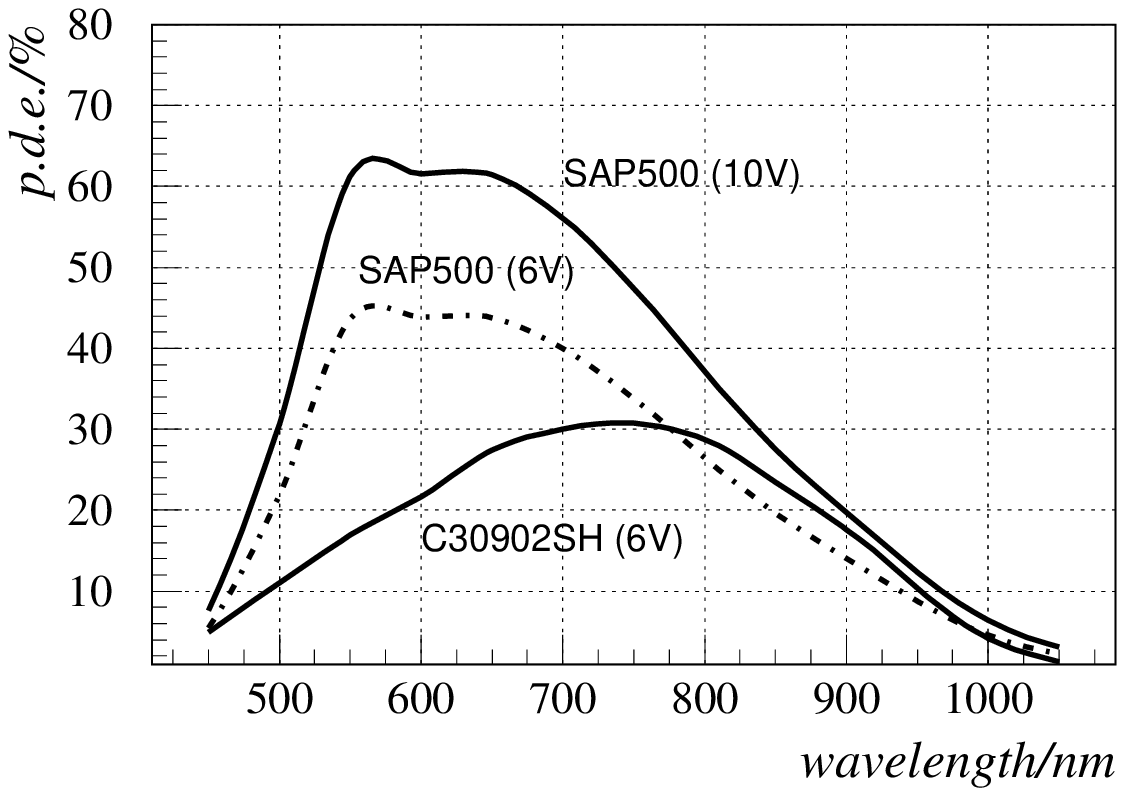}}
\caption{Spectral photon detection efficiency curves obtained at -23.2$^{\rm o}$C: for C30902SH ($V_{over}$=6V, $t_{Q}$=30ns) and for SAP500 ($V_{over}$=6V,10V, $t_{Q}$=9ns).} 
\label{spectral-curves}
\end{figure}

Spectral curves measured for the two SPADs at low temperature (-23.2$^{\rm o}$C) are shown in Fig. \ref{spectral-curves}. The diodes are tested at their maximum overvoltages permitted by our AQ circuit as discussed in the section \ref{undervoltage} (6V overvoltage, $t_Q$=30ns for C30902SH, 10V overvoltage, $t_Q$=9ns for SAP500). For comparison SAP500 has also been tested at 6V overvoltage. 
We se that the peak PDE of C30902SH is at longer wavelength than the peak PDE of SAP500. This is expected since C30902SH has a thicker active region.
On the other hand, due to its lower afterpulsing and smaller undervoltage, SAP500 can
be operated at higher overvoltage using the same AQ circuit, thus effectively offering better PDE at all wavelengths.

\section{Detection efficiency versus overvoltage}

Probably the most important consideration in photon detector design is its detection efficiency which ideally should be as high as possible at all wavelengths of interest. In SPADs, photon detection efficiency, at any chosen wavelength, rises with applied overvoltage and asymptotically approaches quantum efficiency.  

\begin{figure}[h]
\centerline{\includegraphics[width=80 mm,angle=0]{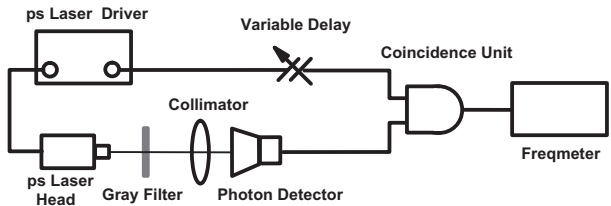}}
\caption{Setup for measuring photon detection efficiency as a function of the overvoltage by use of the pulsed laser and the coincidence technique.} 
\label{devsov-setup}
\end{figure} 

Measurement was performed using the setup shown in Fig. \ref{devsov-setup}. The  picosecond pulsed laser ($\lambda$=676nm, FWHM=39ps) was fired periodically (at a rate of 100kHz) in such a way that either 1 or 0 photons reached the SPAD in every pulse. The coincidence gate is open in synchronization with every emitted photon and thus the frequency meter counts all detected photons while the dark counts are almost completely suppressed due to a very short coincidence window of only 20ns. Furthermore, SPADs were kept at a low temperature (-23.2$^{\rm o}$C) so that the loss of detection efficiency due to dead time caused by dark pulses is kept at a negligible level.
In that conditions, reading at the frequency meter is directly proportional to the photon detection efficiency at 676nm.

\begin{figure}[h]
\centerline{\includegraphics[width=80 mm,angle=0]{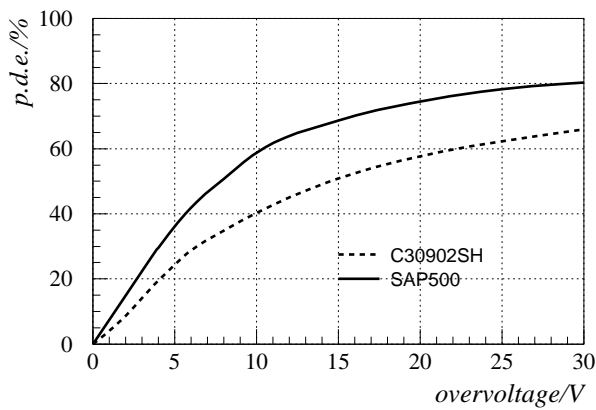}}
\caption{Photon detection efficiencies as a function of the overvoltage at 676nm. SPADs were kept at -23.2$^{\rm o}$C.} 
\label{devsov}
\end{figure} 
 
Knowing the value of p.d.e. at 676nm and at 10V overvoltage (which can be read from the spectral graph in Fig. \ref{spectral-curves}) one can turn these relative efficiencies into absolute ones, which are shown in the Fig. \ref{devsov}. As expected we see that p.d.e. rises with the overvoltage and saturates at some value close to the declared quantum efficiency of each SPAD at the given wavelength. However, as we have seen throughout this paper, the maximum practically achievable p.d.e. depends on levels of noise and afterpulsing probability which also rise with the overvoltage. 

\section{Conclusion}

A new SPAD SAP500 has been evaluated for a single photon detection in the wavelength range of 400-1000nm and compared to the performance of the "standard" C30902SH. SPADs have been operated in the Geiger mode and a wide range of experimental techniques and setups including both passive and active quenching methods was used. We have found that SAP500 driven at 10V overvoltage shows simultaneously an excellent noise performance (down to 50Hz), low afterpulsing ($<$ 0.25\%) and high photon detection efficiency (up to 62\% at 650nm) which makes it an interesting alternative for SPAD based single photon detection applications in VIS-NIR range.

\section{Ackgnowledgements}

This work was supported by project "Photon detector" financed by Croatian Institute of Technology 2007-2010 and by Ministry of science education and sports of Republic of Croatia, contract number 098-0352851-2873. We are indebted to Pavel Trojek from LMU in M\" unchen for his help in assembling the parametric downconversion setup.


\end{document}